\documentclass[prl,twocolumn,showpacs,amsmath,amssymb]{revtex4}

\usepackage{graphicx}
\usepackage{dcolumn}
\usepackage{bm}

\begin{document}

\title{Saffman--Taylor instability in a non-Brownian suspension: \\
finger selection and destabilization}

\author{C. Chevalier}
 \altaffiliation[Present address: ]{Laboratoire Central des Ponts et Chauss\'ees, Paris, France}
\author{A. Lindner}
 \email{lindner@ccr.jussieu.fr}
\author{E. Cl\'ement}

\affiliation{ Laboratoire de Physique et M\'ecanique des Milieux
H\'et\'egog\`enes (PMMH),UMR 7636 CNRS - ESPCI - Universit\'es Paris
6 et 7, 10, rue Vauquelin, 75231 Paris Cedex 05, France }

\date{\today}

\begin{abstract}
We study the Saffman--Taylor instability in a non-Brownian
suspension by injection of air. We find that flow structuration in
the Hele-Shaw cell can be described by an effective viscosity
depending on the volume fraction. When this viscosity is used to
define the control parameter of the instability, the classical
finger selection for Newtonian fluids is recovered. However, this
picture breaks down when the cell thickness is decreased below
approximatively 10 grain sizes. The discrete nature of the grains
plays also a determinant role in the the early destabilization of
the fingers observed. The grains produce a perturbation at the
interface proportional to the grain size and can thus be considered
as a "controlled noise". The finite amplitude instability mechanism
proposed earlier by Bensimon {\it et al.} allows to link this
perturbation to the actual values of the destabilization threshold.
\end{abstract}

\maketitle

{\it Introduction} -- When a low viscosity fluid like air displaces
a viscous, immiscible fluid in a thin channel or Hele-Shaw cell, an
instability develops at the interface, leading to the formation of
fingerlike patterns, called viscous fingers or Saffman--Taylor
instability \cite{saffman1958}. Since the early work of Saffman and
Taylor, this problem has received much attention not only because of
its practical importance but also since it represents an archetype
of many pattern forming systems \cite{Couder1991,
Homsy1987,Bensimon1986}. Driven by practical and fundamental
interests, several viscous fingering studies have lately been
extended to non-Newtonian fluids where a wide variety of strikingly
different patterns are found \cite{McCloud1995}.

Here, we use the Saffman--Taylor instability in a Hele-Shaw channel
as a model system to study the dynamical properties of the interface
between a pure fluid and a non-Brownian suspension. This kind of
particle laden fluid is known to structure under shear as the
particles migrate towards regions of lowest shear and thus change
the local viscosity of the flow. As a consequence, this is a fluid
with complex evolutive properties. A central question is to
understand whether a simple effective continuum description remains
valid and how pattern selection and finger stability can be affected
by the fluid structuration as well as the discrete nature of the
grains.

\begin{figure}
\includegraphics[width=8.5cm]{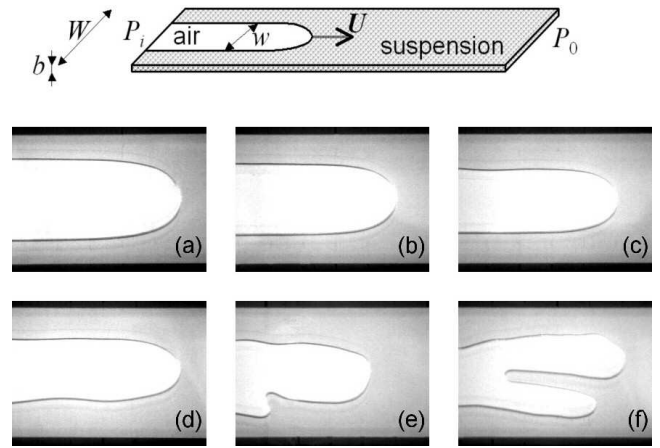}
\caption{Schematic drawing of the experimental set-up. Evolution of
a typical finger with increasing finger velocity
(a)-(f).\label{ChevalierSuspension_1}}
\end{figure}

\begin{figure*}
\includegraphics[width=18cm]{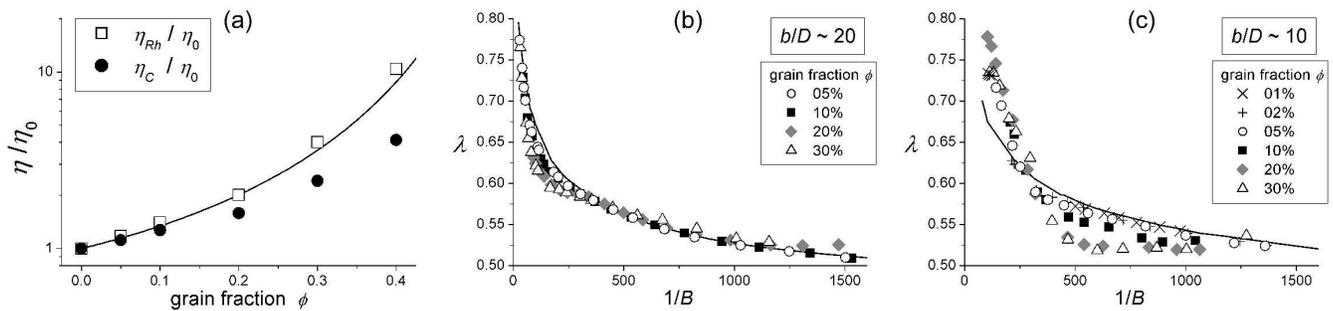}
\caption{\label{ChevalierSuspension_2} a) Relative viscosity of the
suspensions ($D$=80$\mu$m) as a function of the grain fraction
$\phi$ obtained from rheological measurements
$\eta_{Rh}(\phi)/\eta_0$ ($\Box$) and from flow in the Hele-Shaw
cell $\eta_C(\phi)/\eta_0$ ($\bullet$).(--) theoretical prediction
of Zarraga {\it et al.} \cite{Zarraga2000}. b and c) Relative finger
width $\lambda$ as a function of $1/B=12(W/b)^2 (\eta_C U/\gamma)$
for the same suspensions. Cell width $W$=4~cm and cell thickness
$b$=1.43mm (b) and $b$=0.75mm (c)}
\end{figure*}

{\it Set-up and characterization of the suspensions} -- The
experiments are performed in an Hele-Shaw cell of length 1~m formed
by two 1.5~cm thick glass plates separated by a thin mylar spacer
(Fig. \ref{ChevalierSuspension_1}). The cell thickness $b$ can be
varied ($b$=0.75 - 1.43~mm) as well as the width $W$ of the channel
($W$=2 - 4~cm). The thin channel is initially filled with a
non-Brownian suspension that is then displaced by air. The
suspensions are formed by spherical polystyrene beads from Dynoseeds
suspended in a Silicon oil. We use different grain diameters $D$=20,
40, 80 or 140 $\mu$m, their density is $\rho$=1050$\sim$1060
kg/m$^3$. The grains are dispersed in a modified Silicon oil (Shin
Etsu SE KF-6011) such as to obtain density matching at a value
$\rho$=1070 kg/m$^3$. We measured the viscosity of the pure fluid
$\eta_0$ = 191~mPa.s as well as its surface tension $\gamma$ =21$\pm
1$~mN/m at 21$^\circ$C. The fact that the particles and the
suspending fluid are density matched allows us to control the volume
fraction over a wide range. Here we work with volume fractions below
40 $\%$ which allows to avoid jamming problems and wall slip
occurring at higher volume fractions.

A constant overpressure is applied at the inlet whereas the outlet
is maintained at atmospheric pressure. The advancing finger tip is
observed using a CCD camera mounted on a movable system which allows
a manual tracking of its position. The camera is coupled to a
microcomputer for direct image acquisition. Note that in this
configuration, the applied pressure gradient is not constant and
therefore, the finger accelerates when propagating through the cell.
We verified that this acceleration is slow enough and does not
influence the observed finger properties (i.e. for a given tip
velocity, we have no dependence on the applied overpressure).

{\it Finger width selection} -- For Newtonian fluids, the width $w$
of the viscous fingers
 is determined by the capillary
number $Ca= \eta U/\gamma$, the ratio of viscous to capillary
forces; $U$ being the finger velocity, $\eta$ the viscosity and
$\gamma$ the surface tension of the fluid. More precisely, the
relative finger width $\lambda=w/W$ is a function of the
dimensionless control parameter $1/B=12(W/b)^2 Ca$, which involves
the cell aspect ratio.
 The mean flow (averaged over the cell thickness) is
governed by Darcy's law which links the fluid velocity $V$ to the
pressure gradient and far away from the finger, reduces to: $V
=-\frac{b^2}{12 \eta}\nabla P$ where $\nabla P$ is the applied
pressure gradient.

Therefore, by analogy with Newtonian fluids, we seek to establish
the relevant control parameters of the injection process. A first
step will be to measure  the suspension viscosities as a function of
volume fraction. Since  we want to evidence the influence of
structuration effects due to the confinement in the Hele -- Shaw
cell,
 we will compare the viscosities obtained from the commercial rheometer
 to those extracted directly from Darcy's law in the Hele-Shaw cell.

The suspension viscosities  $\eta$ (for $D$=40 and 80~$\mu$m) were
obtained by rheological measurements using a double Couette geometry
rheometer (Haake-RS600) of gap width 2*0.25~mm and mean radius
20~mm. The gap width being small with respect to the radius, the
velocity profiles in the gap can be considered as linear. In the
range of shear rates tested ($\dot{\gamma}$=0.1 - 100 s$^{-1}$)
corresponding also to the typical shear rates of the Hele-Shaw
experiments, the suspensions behave as a Newtonian fluid with a
viscosity $\eta_{Rh}(\phi)$ independent of the grain diameter $D$
and well described by models used in recent literature (as for
example Zarraga {\it et al.} \cite{Zarraga2000}) (see Figure
\ref{ChevalierSuspension_2}a open symbols). These results confirm
the absence of aggregation in the range of volume fractions
considered.

Now we investigate the rheology of our suspensions directly in the
Hele-Shaw cell. To do so, we systematically established a Darcy's
law for all suspensions and cell geometries we have considered. The
details of the procedure can be found in \cite{chevalier2006,
Chevalier2005}. The results indicate the existence of an effective
viscosity parameter specific for flow in a Hele-Shaw cell
$\eta_C(\phi)$. In all the studied configurations, this viscosity
was found to be independent of the cell geometry (i.e. $b$ and $W$)
as well as the grain diameter $D$ and is only function of volume
fraction $\phi$.

We found for increasing $\phi$ an increasing deviation of $\eta_{C}$
from the corresponding rheometer viscosity $\eta_{Rh}$ (figure
\ref{ChevalierSuspension_2} a close symbols). The fact that
$\eta_{C}$ is lower than $\eta_{Rh}$ can  easily be explained by the
effect of flow structuration due to confinement. It is well known
that particles migrate to zones of low shear rate
\cite{Leighton1987}, here in the middle of the gap. Therefore, the
flow profile should deviate from an ideal parabolic profile and
evolve towards a flatter profile \cite{Lyon1997}, independent of $V$
and solely dependent on $\phi$. A steady profil is only reached
after a certain flow distance below which we do not consider our
experiments. In addition we also verified that the grains in the
suspension do not alter the surface tension.

We systematically studied  the selection of the finger width for
stable fingers different values of the volume fraction $\phi$.
Typical results are displayed on figure \ref{ChevalierSuspension_2}b
and c.
 The solid line represents the pure fluid measurements and thus corresponds to the classical
result for Newtonian fluids. Importantly, we choose to use
$\eta_C(\phi)$ to define the control parameter $1/B=12(W/b)^2
(\eta_C U/\gamma)$. We notice that this choice rescales our data
when the ratio between the cell thickness and the grain diameter is
larger then approximatively 10. Below $b/D\sim 10$ significant
deviations from the classical result appear: fingers are slightly
larger for low $1/B$, thinner for intermediate $1/B$ - this
narrowing is more important for large volume fraction  - and then
eventually $\lambda$ will join the classical curve for larger values
of $1/B$. In all cases, for large 1/B the asymptotic value 1/2 for
the relative
 finger width seems to be recovered.

So far, we were not able provide any fully satisfactory argument to
explain the deviation from the classical results. One might however
speculate that effects like a depletion of grains in front of a
receding meniscus \cite{tang2000} (where 3D effects become
important) would lead to a lower viscosity at the finger tip and
thus induce a finger narrowing as observed for shear thinning fluids
\cite{Lindner2002}. One could also consider other effects like
normal stresses existing in granular suspensions \cite{Zarraga2000},
that have been shown to increase the finger width \cite{Lindner2002}
and might thus eventually be responsable for the fact that the
narrow fingers get wider again and stem to the classical results at
a higher control parameter.

{\it Finger destabilization} -- When increasing the finger velocity
and thus the control parameter $1/B$ one observes a destabilization
of the fingers. First, the sides of the fingers start to ondulate
and when the velocity is further increased, the finger gets unstable
by tip splitting. A typical evolution of such a destabilization can
be seen on figure \ref{ChevalierSuspension_1}. Note that in some
cases asymmetric fingers (not shown here) have been observed like
predicted by Ben Amar {\it et al.} \cite{BenAmar1996}.

  \begin{figure}[t]
 \includegraphics[width=8.5 cm]{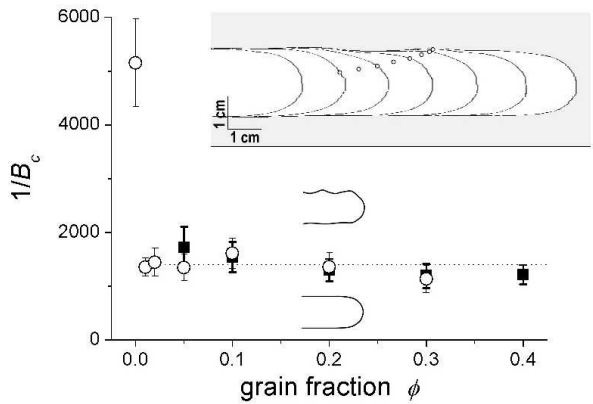}
\caption{Control parameter at the threshold $1/B_c$ as a function of
the grain fraction $\phi$. $W$=4cm, $D$=80$\mu$m and $b$=0.75~mm
($\Box$)
 and $b$=1.43~mm ($\circ$). Inset: Localisation of the perturbation
 ($\circ$)
 during a typical experimental finger destabilization.\label{ChevalierSuspension_3}}
        \end{figure}

This destabilization scenario is also observed in classical
Saffman--Taylor experiments and has so far been attributed to the
noise in the cell hence making it difficult to predict exact values
for this instability threshold \cite{Park1985, Tabeling1987}. In the
case of our suspension, we identified the instability threshold for
different
 volume fractions and cell thicknesses.
 To have a reproducible and objective method of detection,
 we have measured by image processing the fluctuations
of the finger width $\delta\lambda$ and detect when this quantity
starts to increase. The onset of fluctuation growth defines the
stability threshold and therefore, the critical control parameter
$1/B_c$.

When  the values of $1/B_c$ are plotted against the grain fraction
(see figure \ref{ChevalierSuspension_3}) for two different set ups
(different cell thickness, same grain size), we surprisingly find
that as soon as a small amount of grains (as small as 1 $\%$) is
added, the value of the threshold drops strongly and thereafter
varies weakly with the grain fraction (if not constant within the
experimental uncertainties).
 Note that we found different values for the threshold in pure fluid
 for the different
configurations and thus two different intrinsic levels of noise.
However in the presence of grains the value of the threshold is well
defined.

When looking closer at the destabilization scenario of a single
finger (figure \ref{ChevalierSuspension_3} inset), one observes that
perturbations nucleate close to the finger tip and are then advected
to the side of the finger. A possible explanation for this
destabilization mechanism was given by Bensimon {\it et al.}
\cite{Bensimon1986}. They propose that a perturbation would nucleate
at the finger tip, where the normal velocity is the highest and
where the growth rate (given by the linear stability analysis of the
Saffman--Taylor problem \cite{Chuoke1959}) is largest. While the
finger continues to grow, the perturbation is advected to the side
of the finger where the normal velocity goes to zero and thus, the
perturbation growth stops. In the process of advection, the
perturbation is stretched and consequently, its amplitude is
decreased. For a given control parameter $1/B_c$, one needs a given,
finite amplitude of the initial perturbation $A_i$ to be able to
obtain a perturbation with a given final amplitude $A_f$ when the
side of the finger is reached. More quantitatively, Bensimon {\it et
al.} derived the following relation that describe the "finite
amplitude instability":
\begin{equation}
A_f \approx A_i \exp{(0.106\ 1/B_c^{1/2})} \ . \label{BensEq}
\end{equation}

\begin{table}
\caption{\label{tab:config} Critical control parameter $1/B_c$ for
different experimental configurations and $\phi=10\%$.}
\begin{ruledtabular}
\begin {tabular}{c|ccc|c}
  N° config. &  $D$ &  $b$ &  $W$ & $1/B_c$ \\ \hline
  1 & 80 $\mu$m & 0.84 mm & 20 mm & 1210 \\
  2 & 20 $\mu$m & 0.75 mm & 40 mm & 3220 \\
  3 & 40 $\mu$m & 0.75 mm & 40 mm & 2670 \\
  4 & 80 $\mu$m & 0.75 mm & 40 mm & 1600 \\
  5 & 80 $\mu$m & 1.43 mm & 40 mm & 1530 \\
  6 & 140 $\mu$m & 1.43 mm & 40 mm & 1150 \\
  \end{tabular}
  \end{ruledtabular}
\end{table}

\begin{figure}
\includegraphics[width=8.5 cm]{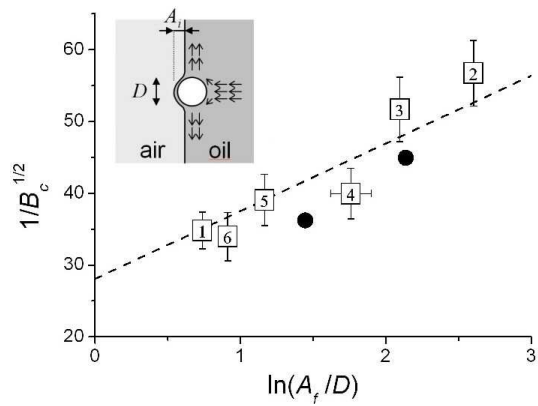}
\caption{\label{ChevalierSuspension_4} Experimental values ($\Box$)
and theoretical prediction form Bensimon {\it et al.} (--). The
numbers correspond to the experimental configuration given in table
\ref{tab:config}. Additional experiments ($\bullet$) using a
different silicon oil (Dow Corning 704).}
\end{figure}

In the following, we will test if the destabilization observed in
our situation can be described by the previous mechanism and if the
existence of grains in the viscous fluid can be directly linked to
the instability onset. A first indication is that we observe a
destabilization threshold hardly affected by an increasing grain
fraction; the above described mechanism is indeed independent of the
wavelength of the perturbation which one might be tempted to link to
the grain fraction. In their theoretical approach, Bensimon {\it et
al} \cite{Bensimon1986} consider the final amplitude $A_f$ to be
proportional to the channel width $W$. For our analysis, we directly
use the value of the fluctuations of the finger width $\delta
\lambda$ observed at $1/B_c$. Those are of course via the finger
width proportional to $W$ and we consider a sinusoidal form of the
fluctuations leading to $A_f=\delta \lambda W 2 \sqrt{2}$. A natural
assumption is to relate the amplitude of the initial perturbation
$A_i$ to the grain diameter $D$: $A_i\propto D$. In table
\ref{tab:config} we sumarize the results obtained for different
grain sizes $D$ and different cell widths $W$. When we
 plot $\sqrt{1/B_c}$ as a function of $\ln(A_f/D)$ obtained for
$\phi=10\%$, different cell geometry and different grain sizes (Fig.
\ref{ChevalierSuspension_4}), we obtain a very satisfying
quantitative agreement with the theoretical result of Bensimon {\it
et al} when the initial perturbation is taken to be: $A_i\approx
D/20$.

Now, to provide a rational justification for our choice for the
initial perturbation, we need to show that a particle approaching a
free interface at a stagnation point flow, as is the case of the
finger tip, would indeed be able to deform this interface with the
correct amplitude. Hoffman \cite{Hoffman1985} and Montiel {\it et
al.} \cite{Montiel2003} established a link between the free
interface amplitude of perturbation and a capillary number defined
on the scale of a particle: $Ca_p=\frac{2}{3}\frac{D}{b}Ca$. In our
case, the range of capillary numbers would be $Ca_p$=5*10$^{-3}$ -
2*10$^{-2}$ and thus, the corresponding values for the amplitude of
the perturbation are of order of $0.02 - 0.1\times D$. Therefore,
this is indeed in good agreement with our choice of the amplitude of
the initial perturbation and it confirms the validity of the
theoretical prediction of Bensimon {\it et al} \cite{Bensimon1986}.
This is a striking emergence of the particulate nature of the
suspension, the grains acting here as a controllable noise
amplitude.

{\it Conclusion} -- We have shown that the characteristics of the
stable Saffman -- Taylor fingers invading a non-Brownian suspension,
are mainly determined by the effective properties of the suspension
in the cell far from the fingers. The structuration effects of the
confined particulate flow can simply  be taken into account by an
effective viscosity, which  allows to rescale the width selection
curve onto the classical results. However, when the grain size
becomes important compared to the cell thickness, systematic
deviations  from the classical result are evidenced.

We also studied the destabilization of the Saffman--Taylor fingers
and we showed that individual grains perturbe the interface between
the air and the suspension and lead to a premature destabilization
of the fingers. The threshold of instability is found to match
quantitatively the theoretical prediction of Bensimon {\it et al.}.
To our knowledge this is the first time that an experiment allows to
control the initial "noise" in the cell and thus to investigate a
mechanism of finite amplitude instability directly.

Recently a number of studies have reported oscillations of the
finger width, observed for example for low capillary number
\cite{moore2002} or fixed perturbations of the cell thickness
\cite{torralba2006}. A more close characterization of the
oscillations resulting from the premature destabilization in our
system might reveal similarities between the different systems.

\begin{acknowledgments}
We wish to thank Michel Clo\^itre and Fabrice Monti for help with
the rheological measurements, Daniel Bonn for for help and advice
with the experimental set-up and Jos\'e Lanuza for technical
assistence.
\end{acknowledgments}

\end{document}